# Superradiant Organic Light-Emitting Diodes

Kieran Hymas[1], Tadahiko Hirai[1], Daniel J. Tibben[2], Jack B. Muir[1], Christopher J. Dunn[1], Daniel E. Gómez[2], James Q. Quach[1]*

1. Commonwealth Scientific and Industrial Research Organisation (CSIRO), Clayton, Victoria 3168, Australia

2. School of Science, RMIT, Melbourne, Victoria 3000, Australia

*Corresponding author: james.quach@csiro.au

**Organic light-emitting diodes (OLEDs) are central to modern display technologies and are promising candidates for low-cost energy-efficient lighting[1,2]. Their performance is determined by numerous, intricate fabrication parameters, but not least by the number of emissive molecules $N$, which provide sites for electron-hole recombination and photon generation in the diode host matrix. Counterintuitively, larger concentrations of emitters do not always lead to brighter or more efficient OLEDs due to concentration quenching of luminescence[3-5] meaning that rates of radiative electron-hole recombination can become severely reduced, negatively impacting charge-to-photon conversion efficiency. In this work we trigger steady-state superradiant light emission from a series of Fabry-Pérot microcavity OLEDs by scaling the operating voltage of each device with emitter concentration. We demonstrate a collective enhancement in the luminance of a microcavity OLED that scales super-extensively when compared to no-cavity controls fabricated in the same run. Triggering quantum correlations between emitters via the confined cavity field allows devices with fewer emitters to match or even exceed the brightness of control OLEDs even when driven by lower voltages. Moreover, our devices show significant narrowing of their emission spectra, offering purer colours at low applied voltages. Leveraging collective effects in microcavity OLEDs provides a new approach to**

**enable brighter, more efficient devices paving the way for next-generation displays and lighting that do not compromise performance for operational efficiency or device lifetime.**

**Main**

Organic light-emitting diodes (OLEDs) have transformed display and lighting technologies, owing to their mechanical flexibility, colour tunability, and compatibility with low-cost, large-area fabrication[6]. Over the past three decades, advances in molecular design, device architecture, and processing techniques have dramatically improved OLED efficiency and performance, establishing them as a cornerstone of consumer electronics and energy-efficient lighting[7]. Given their widespread adoption in modern technologies, pushing OLEDs to their performance limits remains an important but challenging goal, particularly at high emitter concentrations.

In an idealised molecular ensemble under incoherent optical pumping, the stochastic emission of photons is an extensive process that scales linearly with the number of emitters, $N$, since each molecule contributes independently to the total luminescence[8]. Dicke superradiance exemplifies a super-extensive emission process (*i.e.* it scales faster than linear with $N$), in which quantum coherence between emitters leads to a boost in the intensity of the emitted luminance, specifically so that the emitted light intensity scales quadratically with $N$, highlighting the profound impact of many-body correlations on light emission. However, in state-of-the-art OLEDs, photon generation relies on electroluminescence, occurring via radiative recombination of charges that are electrically injected from the electrodes, onto emitter molecules, which are doped into a host matrix in the emissive layer of the device. This basic design is frequently augmented by the addition of charge transporting layers that, together with the host material, facilitate the transport of holes and electrons from thin film electrodes. a process that requires electrical excitation via charge injection and recombination, rather than direct optical pumping.

This fundamental difference introduces a range of concentration-dependent quenching mechanisms that disrupt extensive scaling of the luminescence. For instance, emitter concentration in the emissive layer of Ir(ppy)$_3$-based devices was found to significantly quench radiative recombination beyond just 10% doping owing to clustering of emitters in the host matrix[9] significantly increasing the likelihood of exciton-exciton annihilation, exciton-polaron quenching, and aggregation-induced non-radiative decay[10-13]. Many approaches, including molecular design[14-16], host-guest interface engineering[17,18] and doping optimisation[19], have provided strategies for the mitigation of these phenomena, yet they do not address the intrinsic concentration-dependent limitations imposed by electrical excitation in OLED architectures.

An alternative approach to improve device performance involves triggering super-extensive collective effects (for instance steady-state Dicke superradiance) in the emitter ensemble using optical microcavities[20-24]. This approach can enhance light outcoupling and intrinsically modify the emission characteristics of OLEDs, resulting in increased brightness and directional, narrowband emission[25]. More subtly, so-called polariton-enhanced Purcell effects[1] have already been leveraged to greatly reduce the operational concentration of high energy triplet excitons in blue emitting OLEDs, significantly reducing triplet annihilation events that lead to device degradation[26,27]. Microcavity OLEDs have also been augmented with additional optical layers that exhibit strong light-matter coupling to lessen angular spectral shift[28,29]. Microcavity-induced quantum correlations between emitters in the host matrix present a novel super-extensive resource for controlling exciton dynamics, offering a promising pathway to overcome the intrinsic trade-offs faced by electrically driven molecular emitters. Yet, despite three decades of pioneering research into OLED collective emission, the steady-state superextensive emission of light from microcavity OLEDs has not been experimentally demonstrated.

---

[1] The polariton-enhanced Purcell effect refers to the increased radiative decay rate of triplet excitons in OLEDs due to energy transfer to plasmon–exciton polaritons (PEPs) formed at the metal–dielectric interface.

Here, we demonstrate a mechanism to trigger collective effects in microcavity OLEDs to induce steady-state superradiant emission that overcomes the limitations of high emitter doping, providing a new approach to optimise OLED performance. By coupling the emission to a leaky Fabry-Pérot microcavity, we demonstrate that the steady-state emission scales super-extensively with the number of emitters (with an enhancement $\sim N^{1/4}$), so that devices with fewer emitters, but operating at lower applied voltages, achieve the same brightness as, or even greater brightness than control OLEDs in which collective effects have not been triggered. Moreover, we show that this configuration results in a significant narrowing of the emission spectrum, enabling purer colors at lower operating voltages. Our findings offer new insights into the design of next-generation OLED technologies that combine high performance, operational efficiency, and perhaps, extended device lifetimes.

**Device engineering and steady-state superradiance**

We fabricated six bottom emitting OLED devices (D1 to D6) containing differing concentrations of the phosphorescent emitter Ir(ppy)$_3$ ranging from 5% to 55% v/v doping in a 1,3-bis(N-carbazolyl)benzene (mCP) host matrix (see Supplementary Table 1). Each device was fabricated with six spatially separated pixels deposited simultaneously on a shared ITO-glass substrate in the same fabrication run. An annotated schematic of the OLED stack is shown in Figure 1a (where the bottom emitting stack is show upside-down) where the thick Al and semi-transparent Ag mirrors form a Fabry-Pérot microcavity architecture encasing the Ir(ppy)$_3$ emitters and other charge transport layers (CTLs) and charge injection materials. We incorporated charge transport layers with HOMO-LUMO bandgaps (Figure 1d) into the microcavity to effect an energy gradient that favoured efficient electron and hole diffusion to the Ir(ppy)$_3$ doped emissive layer. Each pixel strongly emitted green light when a voltage was applied to the ITO layer shared by all pixels and the Al mirror (see Figure 1b). One of the six pixels was fabricated without the semi-transparent Ag mirror, acting as a "no-cavity" control for each device. The emission spectrum of the control device, plotted in Figure 1c, shows a broad emission centred about 530 nm and agrees well with photoluminescence and electroluminescence measurements on similar Ir(ppy)$_3$ thin film devices[30].

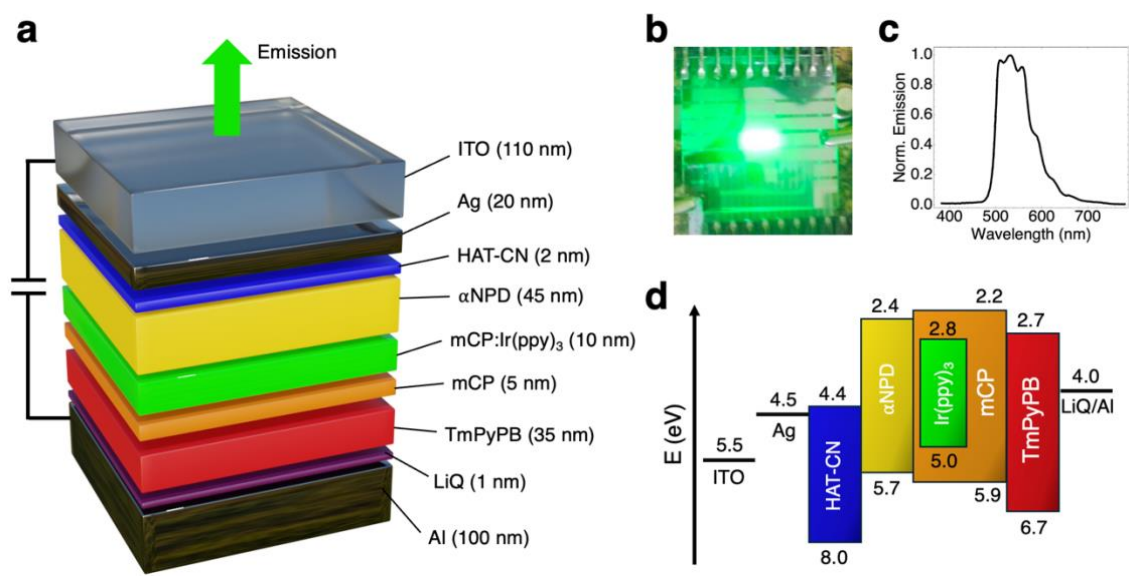

**Figure 1:** Design of a microcavity-coupled OLED. **a** Schematic depiction of the OLED stack with false colouring. The emitters Ir(ppy)$_3$ are doped in a mCP host layer flanked by two optical mirrors comprised of Al and Ag forming a microcavity. The Al mirror also acts as an electrical contact with an ITO layer outside the microcavity which facilitates electrical pumping of the Ir(ppy)$_3$ excitations. **b** Light emission from one pixel of the microcavity OLED when a voltage is applied. **c** Normalised emission spectrum of Ir(ppy)$_3$ from a control OLED device missing the top Ag mirror. **d** Energy level diagram of the layers of the OLED stack indicating the work functions of the ITO, Ag and LiQ-dressed Al layers and the HOMO-LUMO energy levels of all other materials.

We modelled the photon flux emanating from the microcavity OLED using an incoherently driven Tavis-Cummings model[31,32] (see Methods and Supplementary Note 2) where the incoherent drive $\omega(V)$ represents the rate of excitations, induced by the application of a voltage V across the electrodes of the device, to the spin-orbit coupled Ir(ppy)$_3$ excited state. The analytical solution of the coupled equations of motion indicated that the photon flux from the microcavity OLED benefits from collective enhancements that scale as $\sim N^2 \langle \sigma^+ \sigma^- \rangle$ where $N$ is the number of emitters coupled to the fundamental mode of the cavity photon field and $\langle \sigma^+ \sigma^- \rangle$ is the same-time two-body correlator between two distinct emitters in the ensemble. Crucially, we found that $\langle \sigma^+ \sigma^- \rangle \sim \omega/N$, so that the collective enhancement of the OLED photon flux as a function of dopant concentration is achieved only when the voltage-mediated incoherent drive $\omega(V)$ is adequately scaled with $N$.

In Figure 2a we plot the ratio of the luminance from the cavity $L_c$ and no cavity control $L_{nc}$ pixels of each device while the voltage is varied linearly according to the emitter concentration (inset). In the absence of collective effects, the photon flux of a stochastically emitting ensemble should scale linearly with $N$ so that a scaling $L_c/L_{nc} \sim N^\alpha$ with $\alpha > 0$ is indicative of a collective enhancement of light outcoupling in the microcavity devices. Examining the ratio $L_c/L_{nc}$ rather than $L_c$ directly controls for fabrication variability with changing $N$, since the cavity and control pixels of each device were deposited in the same fabrication run. We show that, within error, the luminance from our OLED devices scales as $L_c/L_{nc} \sim N^{1/4}$, within the bounds predicted by theory $0 \leq \alpha \leq 1$: this means that for every microcavity device, the light emission increases faster than linear with $N$. As shown in Supplementary Note 2, achieving the theoretical maximum collective enhancement would require much higher voltages and current densities than are currently possible with our prototype devices, nevertheless this bound sets an important milestone for future devices that may involve different materials with more intricate stack designs.

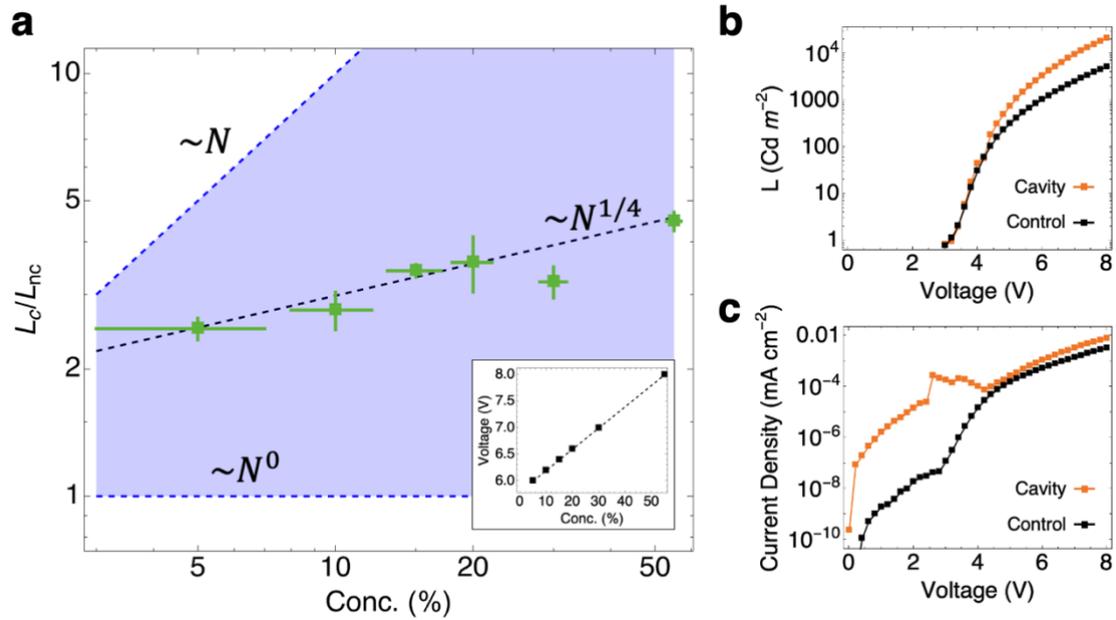

**Figure 2:** A steady-state superradiant OLED. **a** Ratio of the cavity versus no-cavity luminance for each device (green squares) as the voltage applied to each device is scaled linearly with concentration (inset). Vertical error bars indicate the first standard deviation of the luminance ratio measurement and horizontal error bars indicate uncertainties in the doping concentration. As the dopant concentration increases, the luminance ratio scales as $N^{1/4}$, indicated by the dashed black line fitted to the data with RMSD 0.12. This scaling falls within the theoretical bounds (shading blue region) of optimal steady-state superradiance (where $L_c/L_{nc} \sim N$) and no collective enhancement (where $L_c/L_{nc} \sim N^0$), shown as dashed blue lines. **b** Luminance and **c** current density versus applied voltage for the microcavity (orange) and the control (black) pixels of D4 (20% Ir(ppy)$_3$ dopant concentration). Squares are experimental data points and solid lines are guides for the eye.

In Figure 2b and 2c we show the luminance and current density, respectively, versus voltage for a representative device (D4). The luminance enhancement of the cavity versus control is particularly apparent at large, applied voltages where a slightly higher current density flows through the microcavity device. The average charge-to-light conversion efficiency of all six devices was 46.8 Cd A$^{-1}$ when tuned to a luminance of 10$^3$ Cd m$^{-2}$. Comparatively, the controls reached a mean efficiency of 36.8 Cd A$^{-1}$ under the same conditions.

**Narrowband emission**

We next characterised the emission properties of the microcavity OLED devices compared to their control counterparts. In Figure 3a we compare the emission spectrum between the microcavity (orange) and control (black) pixels of D4. The microcavity electroluminescence spectrum is visibly narrower (30 nm FWHM) giving a much purer green colour at approximately 527 nm compared with the broad emission from the control. The colours emitted from the cavity and control pixels are quantified in the inset using the CIE 1931 colour space chart. The average FWHM for our six devices was 35 nm exhibiting a large improvement over the ~75 nm FWHM of the controls. A narrow OLED spectral emission would remove the necessity for colour filters that are currently used in state-of-the-art OLED displays to achieve high colour purity[33]. The removal of these filters represents a further gain in efficiency for displays based on microcavity OLEDs.

Angle-resolved electroluminescence measurements (Figure 3b) for the control and cavity pixels of D4 highlight the enhanced emission of the microcavity, which emits nearly twice as many photons as the control pixel over the same detector exposure time. The microcavity pixel exhibits a slight angle dependence in its emission frequency, a typical signature of emitter and cavity mode hybridisation[29]. This angle dependence can be mitigated by engineering the detuning between the cavity mode and the emitter frequency[28]. Alternatively, the angle-dependent emission could be leveraged as a functional feature, for example in privacy screen applications. Angle-dependent emission is furthermore a signature of coherent light emission, which is predicted from the synchronised radiative-relaxation of emitters[34,35]. In Supplementary Note 2 we also present calculations from our theoretical model of the Hanbury-Brown-Twiss signal that shows the potential for coherent emission from a driven microcavity OLED.

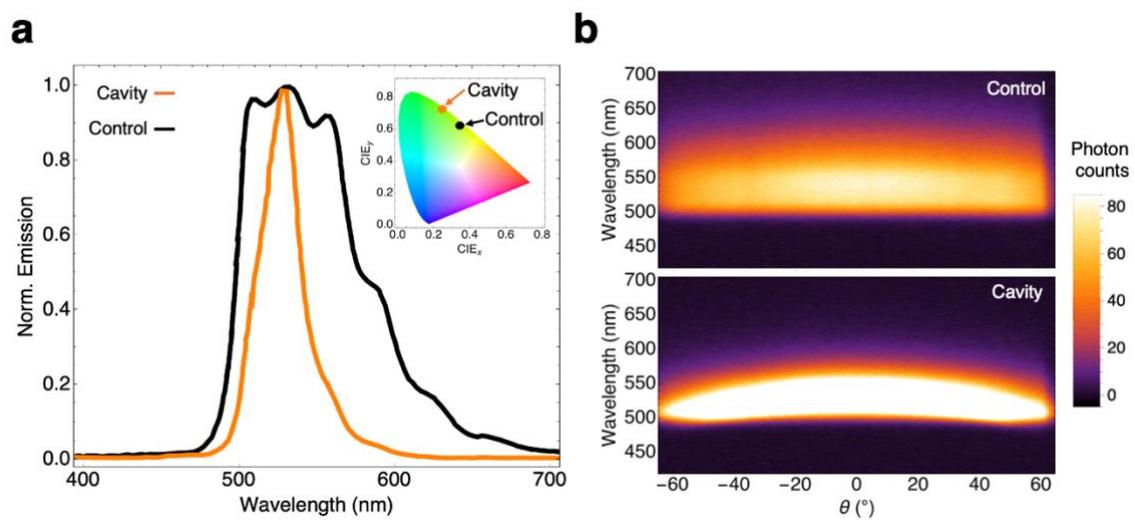

**Figure 3:** Narrowband angular-resolved emission. **a** Electroluminescence spectra of the microcavity (orange) and control (black) pixels of D4 with an applied voltage of 8 V. The microcavity gives a narrow emission peak at 527 nm with a FWHM of 30 nm. The colours of light emitted from the microcavity and control pixels in D4 are quantified using the CIE colour space (inset). **b** Angle-resolved electroluminescence photon counts for the control (top) and microcavity (bottom) pixels of D4 with an applied voltage of 6V and an exposure time of 0.1 s. The microcavity pixel exhibits angle-dependent emission as a consequence of coupling the electrically stimulated emitters to the angle-dependent cavity mode.

**Angle-resolved Reflectometry**

To gain deeper insight into the light–matter coupling in our OLED devices, we performed angle-resolved reflectometry on the cavity pixels and compared the resultant spectra with predictions from our theoretical model. A strong agreement between the experimental and theoretical reflectance spectra is shown in Figure 4 for the cavity pixel of device D4. The theoretical spectrum was calculated using realistic parameters—including the emitter transition frequency $\Delta = 2.35$ eV for Ir(ppy)$_3$ and angular dispersion $\Delta_c(\theta)$ (both overlaid)—and shows quantitative agreement with the experimental data. From this model, we extracted a collective light–matter coupling strength $g = 11$ meV for the microcavity of D4. Using our theoretical model, we reproduced the microcavity reflectance spectra of all other devices (Supplementary Figure 10) and found the collective light-matter coupling strength scaled with emitter number as $\sqrt{N}$.

In a previously reported microcavity OLED[28], a sizable Rabi splitting (~140 meV) was observed in reflectance measurements at ~20° incidence. Similarly, our model predicts an avoided crossing between the coupled oscillator eigenstates near 20°, giving rise to polaritonic branches with a Rabi splitting of ~20 meV. However, this splitting is not resolved in the reflectance measurements due to homogeneous broadening of the reflectance lineshape; this is expected for the fast (leaky) cavity regime employed here to achieve super-extensive light emission scaling. In contrast, we observed two distinct, angle-dependent polaritonic branches in the emission spectrum of an analogous cavity device with a thicker semi-transparent Ag mirror (Supplementary Note 4), which likely operates in a more strongly coupled, less leaky regime.

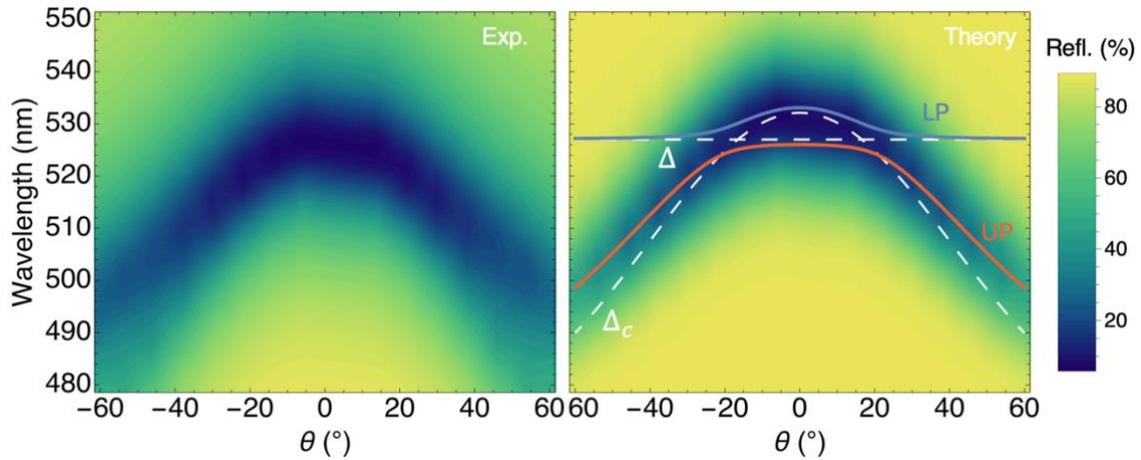

**Figure 4:** Angle-resolved reflectometry. Comparison of the angle-resolved reflection spectrum of the microcavity pixel of D4 with theoretical spectrum using a coupled oscillator model. The angle-dependent cavity mode $\mathbf{\Delta_c}$ and emitter frequency $\mathbf{\Delta}$ used in the coupled oscillator model are plotted as dashed white lines and the lower (LP) and upper (UP) polariton branches are plotted as solid blue and red lines, respectively.

**Discussion**

Engineering quantum collective effects in OLED devices offers a viable and immediate strategy to improve charge-to-photon conversion efficiency, combating efficiency roll-off at high brightness, a long-standing bottleneck in OLED performance for commercial displays and solid-state lighting[36]. While conventional strategies to improve OLED efficiency, such as molecular design for enhanced exciton formation and triplet harvesting, have been highly successful, they do not introduce novel physical mechanisms to enhance light outcoupling at lower applied voltages. In contrast, the electrically stimulated cooperative emission described here amplifies radiative rates super-extensively when the applied voltage is also scaled with emitter concentration, thus providing a fundamentally new lever to boost device performance. Importantly, the voltage-tuned collective emission enhancements demonstrated in our device are complementary to established synthetic and structural strategies—such as emitter molecular design and device architecture optimisation—that improve charge-to-photon conversion efficiency in OLED technologies.

Additionally, our microcavity OLED devices show directional, narrowband emission—distinct signatures of a coherent light source. The conversion of charge to coherent light is of particular interest for emerging applications in holographic displays and holographic microscopy, where controlled phase fronts and high spatial coherence are critical for accurate image reconstruction[37-39]. The integration of superradiant OLEDs in such technologies could provide thin, flexible, and energy-efficient coherent light sources while maintaining high visual fidelity. The coherence length of light emitted from our OLED devices can be approximated from Figure 3a as $L_{coh} \approx \frac{\lambda^2}{\Delta\lambda} \lesssim 10~\mu m$. While this short coherence length is currently prohibitive for commercial holographic displays, it is consistent with the coherence lengths required for holographic microscopy[40-42]. With continued materials optimisation and improved tolerance to higher driving voltages, it is entirely feasible that the spectral FWHM, $\Delta\lambda$, can be further narrowed—paving the way for practical implementation of our OLED devices in next-generation commercial holographic displays.

Beyond technological implications, we propose OLED stacks as a practical and tuneable platform to explore electrically-driven super-extensive physics that can offer new insights into cooperative emission and non-equilibrium charge dynamics in a room-temperature, solid-state quantum system. While this work has focused on steady-state superradiant

emission—arguably the most pragmatic regime for immediate technological impact—the detailed time-evolution of our devices under time-dependent electrical driving remains unexplored. The application of pulsed or modulated currents may induce new operational regimes and shed light on the dynamical interplay between charge transport and collective light emission, potentially revealing non-Markovian effects arising from memory in the electronic reservoirs and their back-action on the light–matter coupling[43,44]. In this way, superradiant OLEDs not only advance next-generation optoelectronic technologies, but also open a compelling testbed for probing super-extensive optical phenomena using an electric current.

## Materials and Methods

### Device fabrication

ITO-coated glass substrates (Asahi Glass Co. Ltd., Japan) were first cleaned using isopropanol alcohol and deionized water in an ultrasonic bath. The substrates were then treated with UV-ozone in a Novascan PSD Pro cleaner to create an oxygen-rich ITO surface, enhancing its work function. Next, a 15 nm thick silver layer (Ag, Sigma-Aldrich Solutions) was deposited onto the ITO, followed by a 2 nm hole-injection layer of 1,4,5,8,9,11-Hexaazatriphenylenehexacarbonitrile (HAT-CN, Luminescence Technology Co.) under high vacuum. A 45 nm layer of the hole transport material N,N′-bis(naphthalen-1-yl)-N,N′-bis(phenyl)-2,2′-dimethylbenzidine ($\alpha$NPD) was then deposited by thermal evaporation under high vacuum ($<1.0\times10^{-4}$ Pa). A 10 nm 1,3-bis(N-carbazolyl)benzene (mCP) host layer was subsequently deposited with varying levels of doping (5%, 10%, 15%, 20%, 30%, and 55% v/v) of the emissive molecule tris(2-phenylpyridine)iridium(III) (Ir(ppy)$_3$). A neat 15 nm mCP layer was then followed by a 35 nm layer of the electron transport material 1,3,5-tris(3-pyridyl-3-phenyl)benzene (TmPyPB). Finally, a 1 nm 8-Quinolinolato lithium (LiQ) electron injection layer was deposited, and a 100 nm aluminium (Al) top mirror layer was added to complete the device stack. The devices were encapsulated under a $N_2$ atmosphere with cover glass and a desiccant, then sealed with UV-curable epoxy resin. The final device dimensions are approximately 0.5 × 2 cm², with four 2 × 5 mm² regions of cavity and two 2 × 5 mm² regions without cavity.

**Electroluminescence and current measurements**

The OLED devices were analysed by measuring their current-voltage and luminescence characteristics using a Keithley 2400 source measure unit and a Topcon BM-7 luminance colorimeter controlled by a computer, respectively. The electroluminescence (EL) spectra were recorded with an ASEQ Instruments LR1 fibre-optic spectrometer with the tip of the fibre placed directly over the centre of each pixel.

Angle-resolved electroluminescence in the devices was stimulated using a Keithley 2401 SourceMeter at 6 V and 2 mA compliance. The resulting emission was collected at the back-focal plane (100x 0.9 NA optical air objective) through a Bertrand lens to resolve angular emission up to 64° from the substrate normal. Spectra were resolved at 0.1 s exposure intervals using an Andor SR-303i spectrophotometer with Andor iDus-420A CCD detector.

**Angle-resolved reflectance**

Angle–resolved reflectometry measurements were performed with an Agilent Cary 7000 UV-Visible-NIR spectrophotometer with Universal Measurement Accessory (UMA) and xenon lamp source. Measurements across the devices were baseline-corrected and performed with standardised spot size and lamp intensity. Fixed-angle reflectometry and transmission measurements were performed with an Agilent Cary 5000 UV-Visible-NIR spectrophotometer with Diffuse Reflectance Accessory (DRA) and xenon lamp source. Measurements across the devices were baseline-corrected and performed with standardised spot size and lamp intensity. Reflectance measurements were collected at 8° angle of incidence to enable collection of specular and diffuse reflectance signal.

**Theoretical model of steady-state photon emission**

To model the steady-state superradiant emission of our devices, we employ the Tavis-Cummings model

$$H = \Delta_c a^\dagger a + \sum_{n=1}^{N} \left( \Delta \sigma_n^+ \sigma_n^- + g(a^\dagger \sigma_n^- + \sigma_n^+ a) \right)$$

where $a^\dagger$ and $a$ are creation and annihilation operators for the photon field confined to the microcavity with fundamental frequency $\Delta_c$. The $\sigma_n^+$ and $\sigma_n^-$ are excitation and de-excitation operators for molecule $n$ in the Ir(ppy)₃ ensemble. The light-matter coupling between the confined photon field and excitations in the Ir(ppy)₃ ensemble is quantified by $g$.

The interaction of the emitters with their environment is described within a Lindblad framework where the evolution of the density operator is given by

$$\dot{\rho} = -i[H, \rho] + \kappa \mathcal{L}[a] + \sum_{n=1}^{N} (\omega(V)\mathcal{L}[\sigma_n^+] + \gamma^- \mathcal{L}[\sigma_n^-] + \gamma^z \mathcal{L}[\sigma_n^z])$$

where the Lindblad superoperator $\mathcal{L}[a] = a\rho a^\dagger - \frac{1}{2}\{a^\dagger a, \rho\}$ describes photon loss from the cavity at a rate $\kappa$. The emitters are incoherently, electrically pumped to the excited state at a voltage-dependent rate $\omega(V)$, non-radiatively relax with rate $\gamma^-$ and lose phase coherence at a rate $\gamma^z$ (see Supplementary Note 2 for more details).

The microcavity OLED luminance under electrical driving is proportional to $L = \kappa \langle a^\dagger a \rangle_{SS}$ where $\langle ... \rangle_{SS}$ denotes the expectation value at steady-state. To obtain this quantity, we numerically solve the coupled equations of motion for the operators $\sigma^z$, $\sigma^+\sigma^-$ and $\sigma^z\sigma^z$ which encode the excited state population of the molecular subsystem, photon field-induced two-body correlations and phase coherence, respectively. The equations of motion for the two-body operators (e.g. $\sigma^+\sigma^-$) contain three-body terms that we approximate as $\langle ABC \rangle \approx \langle A \rangle \langle BC \rangle + \langle B \rangle \langle AC \rangle + \langle AB \rangle \langle C \rangle - \langle A \rangle \langle B \rangle \langle C \rangle$ closing the hierarchy of coupled equations.

**Acknowledgements**

K. H. and J. Q. Q acknowledges funding from the Revolutionary Energy Storage Systems Future Science Platform.